\documentclass{Rinton-P9x6}

\begin{document}

\title{Quantum fluctuations in superresolving  microscopy with squeezed light}

\author{Vladislav N. Beskrovnyy and Mikhail I. Kolobov}

\address{Laboratoire PhLAM, Universit\'e de Lille 1,\\
F-59655 Villeneuve d'Ascq Cedex, France\\
E-mail: mikhail.kolobov@univ-lille1.fr}

\maketitle

\abstracts{ We numerically investigate the role of quantum
fluctuations in superresolution of optical objects. First, we
confirm that when quantum fluctuations are not taken into account,
one can easily improve the resolution  by one order of magnitude
beyond the diffraction limit. Then we investigate the standard
quantum limit of superresolution which is achieved for
illumination of an object by a light wave in a coherent state. We
demonstrate that this limit can be beyond the diffraction limit.
Finally, we show that further improvement of superresolution
beyond the standard quantum limit is possible using the object
illumination by a multimode squeezed light.}

Last years have witnessed an increasing interest to investigation of
quantum effects in optical imaging~\cite{kolobov99,EPJD2003}. One of the
questions recently reconsidered in the light of this latest development is
about the ultimate quantum limits of resolution in optical systems. A
classical resolution criterion formulated at the end of the last century
by Abbe and Rayleigh states that the optical resolution is limited by
diffraction present in any optical system due to the wave nature of light.
This diffraction limit was introduced by Rayleigh for a simple observation
of a diffracted image by a human eye. However, nowadays using the modern
CCD cameras for detection of optical images with subsequent electronic
treatment of the digitized signals one can often improve the resolution
beyond the limit imposed by diffraction. Such superresolution techniques
use some a priori information about the input object and are limited not
by diffraction but by different kinds of noise in the detection and the
electronic reconstruction systems. It was recently
demonstrated~\cite{kolobov00} that the ultimate limit of superresolution
is determined by quantum fluctuations of light, and that the use of
special kind of spatially multimode squeezed light should allow to
increase the capability of superresolution schemes.

\begin{figure}[h]
\begin{center}
\epsfxsize=15pc %
\epsfbox{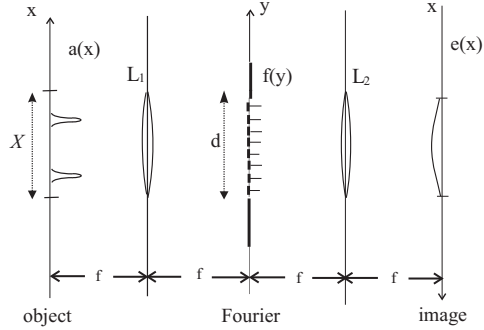}
\end{center} %
\caption{Optical scheme of one-dimensioned diffraction-limited
optical imaging. %
\label{fig:scheme}}
\end{figure}

In this paper we numerically simulate the role of quantum fluctuations for
superresolution of two simple optical objects placed close to each other
so that they cannot be resolved according to the Rayleigh criterion. We
consider a simple one-dimensional scheme of diffraction-limited coherent
optical imaging shown in Fig.~\ref{fig:scheme}. The object of finite size
$X$ is situated in the object plane. The first lens $L_1$ performs the
Fourier transform of the object into the Fourier plane where a pupil of
finite size $d$ is located. Diffraction on this pupil is a physical origin
of the finite resolution distance in the scheme. The second lens  $L_2$
performs the inverse Fourier transform and creates a diffraction-limited
image in the image plane.
As mentioned above, to achieve superresolution one needs some a priori
information about the object. In our case we know a priori that the object
is confined within the area of size $X$ and is identically zero outside.
The spatial Fourier transform of such an object is an entire analytical
function. Therefore, knowing the part of the Fourier spectrum within the
area $d$ of the pupil allows for an analytical continuation of the total
spectrum and, therefore, for unlimited resolution. However, this
analytical continuation is extremely sensitive to the noise in the
diffracted image and, as will be illustrated below, is limited by the
quantum fluctuations of light.

To simplify notations we shall use the dimensionless coordinates $s= 2x /
X$ in the object and the image planes and the dimensionless coordinate
$\xi= 2y / d$ in the pupil (Fourier) plane. Let the classical
dimensionless complex amplitudes of the electromagnetic field in  the
object, Fourier, and the image planes be respectively $a(s)$, $f(\xi)$,
and $e(s)$. The amplitudes $a(s)$ and $f(\xi)$ are related by the Fourier
transform performed by the lens $L_1$,
\begin{equation}
     f(\xi)=\sqrt{\frac{c}{2 \pi}}\int_{-1}^1  a(s) e^{i c s \xi} ds,
     \label{eq:fourier}
\end{equation}
where $c=\pi d X / (2 \lambda f)$ is the space-bandwidth product of the
imaging system.

The second lens $L_2$ performs the inverse Fourier transform and creates
an image in the image plane. The object and the image complex amplitudes
are related by an integral operator,
\begin{equation}
     e(s)=\int_{-1}^1 %
     \frac {\sin [c(s - s^\prime)]}{\pi (s - s^\prime)} a(s^\prime)
     ds^\prime.
     \label{eq:object-image}
\end{equation}

The orthonormal eigenfunctions of this operator are given by
\begin{equation}
     \varphi_k(s)=%
     \frac{1}{\sqrt{ \lambda_k}} \psi_k(s), \quad |s| \le 1,
     \label{eq:eigenfuns}
\end{equation}
where $\psi_k(s)$ are the prolate spheroidal functions and $\lambda_k$ are
the corresponding eigenvalues\cite{slepian61}. To achieve superresolution
one can decompose the object field in the basis of the eigenfunctions
$\varphi_k(s)$ as
\begin{equation}
     a(s)=%
     \sum_{k=0}^{\infty} a_k \varphi_k(s),  \quad |s| \le 1,
     \label{eq:sum-obj}
\end{equation}
with the coefficients $a_k$ calculated by
\begin{equation}
     a_k=\int_{-1}^{1} a(s) \varphi_k(s) ds. %
     \label{eq:decomp-obj}
\end{equation}
Similar decomposition can be written in the image plane,
\begin{equation}
     e(s)=%
     \sum_{k=0}^{\infty} e_k \psi_k(s),  \quad -\infty <s< \infty,
     \label{eq:decomp-four}
\end{equation}
and in the Fourier plane,
\begin{equation}
     f(\xi)=%
     \sum_{k=0}^{\infty} f_k \varphi_k(\xi),  \quad |\xi| \le 1.
     \label{eq:sum-four}
\end{equation}
The coefficients $e_k$ are calculated as
\begin{equation}
    e_k = \int_{-\infty}^{\infty} e(s) \psi_k (s) d s,
    \label{eq:sum-img}
\end{equation}
while the coefficients $f_k$ are given by
     \begin{equation}
     f_k = \int_{-1}^{1} f(\xi) \varphi_k (\xi) d \xi.
     \label{eq:decomp-img}
\end{equation}
Using the properties of the prolate functions $\psi_k$ it can be shown
that the coefficients $e_k$ and $f_k$ are expressed through $a_k$ as
follows:
\begin{equation}
     e_k= \sqrt{\lambda_k} a_k, %
     \label{eq:trans-four}
\end{equation}
\begin{equation}
     f_k= i^k \sqrt{\lambda_k} a_k. %
     \label{eq:trans-img}
\end{equation}
Therefore, detecting the image $e(s)$ in the image plane using, for
example, a sensitive CCD camera, one can calculate the coefficients $e_k$
according to (\ref{eq:sum-img}) and than reconstruct exactly the
coefficients $a_k$ of the object using (\ref{eq:trans-four}).
Alternatively, one can set the CCD camera in the Fourier plane to detect
$f(\xi),$ evaluate the coefficients $f_k$ according to
(\ref{eq:decomp-img}) and reconstruct $a_k$ using (\ref{eq:trans-img}).
The first method could be called {\it superresolving microscopy}, while
the second one {\it superresolving Fourier-microscopy} since one detects
the Fourier spectrum. It should be noted that, since in both cases we need
the complex field amplitudes and not the intensities, one should use the
homodyne detection scheme with a local oscillator.

In our numerical simulations we have tried both detection schemes and have
given preference to the Fourier-microscopy since it involves integration
over the finite region of the pupil, while detecting the image $e(s)$
requires integration over an infinite area in the image plane. It turns
out that due to oscillating behavior of the prolate functions $\psi_k(s)$
one needs to take unrealistically large area in the image plane to achieve
significant superresolution.

\begin{figure}[h]
\begin{center}%
\epsfxsize12.5cm
\epsfbox{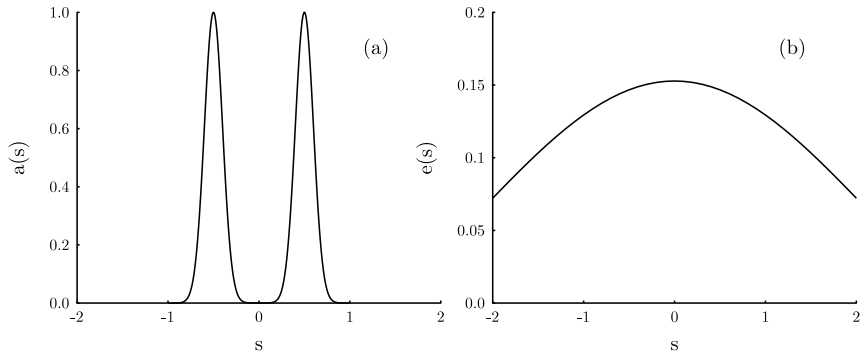}%
\end{center}
\caption{Double-peak object $a(s)$  used in numerical
simulations (a) and its image $e(s)$ (b).%
\label{fig:obj-img}}
\end{figure}

For numerical simulations we have taken a simple object of two Gaussian
peaks,
\begin{equation}
     a(s)=  \exp{\left(-\frac{(s-s_0)^2}{2\sigma^2}\right)} +
     \exp{\left(-\frac{(s+s_0)^2}{2\sigma^2}\right)},  \quad |s|\le 1,
     \label{eq:class-object}
\end{equation}
of width $\sigma$ separated by distance $2s_0$. We choose $2s_0=1$ and
$\sigma=0.1$, so that two peaks are well separated in the input object.
The Rayleigh resolution distance $R=\pi X/(2c)$ in dimensionless
coordinates is equal to $\pi/c$, where $c$ is the space-bandwidth product.
In our simulations we work with $c=1$. In this situation for $2s_0<\pi$ we
are beyond the Rayleigh limit. This is clearly seen in
Fig.~\ref{fig:obj-img} where we have shown the input object and its image
observed in the image plane.

\begin{figure}[h]
\begin{center}
\epsfxsize=14pc %
\rotatebox{270}{\epsfbox{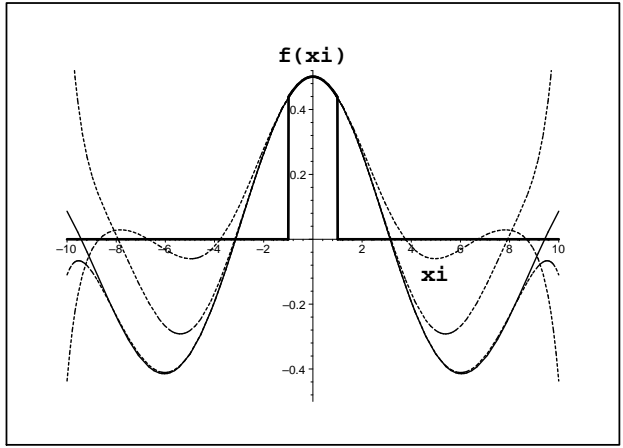}}
\end{center} %
\caption{Spatial Fourier spectrum of the object from Fig.~2a (solid line),
its part transmitted trough the pupil (bold solid line), and the spectra
reconstructed
with 2, 4, and 6 prolate functions (three dotted bold lines).%
\label{fig:spectra}}
\end{figure}

Therefore, observing the image in Fig.~\ref{fig:obj-img}b, it is
impossible according to the Rayleigh criterion to resolve two Gaussian
peaks in input object. However, applying the superresolution technique one
can easily reconstruct the input object beyond the diffraction limit. We
illustrate the result of such a reconstruction in Fig.~\ref{fig:spectra}.
In this figure we show the exact Fourier spectrum of the input object,
drown by a solid line, as a function of dimensionless coordinate $\xi$ in
the Fourier plane. Only a part of this spectrum shown by a bold line,
within the area of the pupil, $|\xi| \le 1$, is transmitted to the image
plane. This is a reason of very large diffraction spread in the image
plane shown in Fig.~\ref{fig:obj-img}b. Three dotted bold lines in
Fig.~\ref{fig:spectra} correspond to the Fourier spectrum of the
reconstructed object using 2, 4, and 6 prolate functions. We can see that
the reconstructed spectrum approaches the exact one for ever higher
spatial frequencies $|\xi|$ as the number of prolate functions increases.
With 6 prolate functions two spectra are very close to each other for
spatial frequencies $|\xi| \le 8$. This corresponds to a superresolution
factor of 8 over the Rayleigh limit.

Up to now we did not take into account the fluctuations in the detection
of the Fourier components by a CCD camera in the Fourier plane. However,
such fluctuations are always present in the detection scheme due to
technical imperfections and the quantum nature of light. The quantum
fluctuations of light set the ultimate limit of superresolution in optical
imaging. The quantum theory of the optical imaging scheme in Fig.~1 was
developed in Ref.~[3]. Applying the same methods for the
Fourier-microscopy, we can write the photon annihilation operators $\hat
a(s)$ and $\hat f(\xi)$ in the object and the Fourier plane as
\begin{equation}
     \hat a(s) = \sum_{k=0}^\infty \hat a_k \varphi_k(s) +
     \sum_{k=0}^\infty \hat b_k \chi_k(s), %
     \label{eq:quant-presnt-obj}%
\end{equation}
\begin{equation}
     \hat f(\xi) = \sum_{k=0}^\infty \hat f_k \varphi_k(\xi) +
     \sum_{k=0}^\infty \hat g_k \chi_k(\xi). %
     \label{eq:quant-presnt-img}%
\end{equation}

Here $\chi_k$ are the orthonormal basis functions in the region $|s|>1$
and $|\xi|>1$ introduced in Ref.~[3], and $\hat b_k$ and $\hat g_k$ are
the corresponding annihilation operators. It can be shown that the
operator-valued Fourier coefficients $\hat f_k$ are given by
\begin{equation}
     \hat f_k = %
     i^k(\sqrt{\lambda_k} \hat a_k + \sqrt{1-\lambda_k} \hat b_k ).
     \label{eq:quant-trans}%
\end{equation}
This relation is similar to the transformation performed by a
beam-splitter with amplitude transmission coefficients
$i^k\sqrt{\lambda_k}$ and reflection coefficients $i^k\sqrt{1-\lambda_k}$,
and preserves the commutation relation of the annihilation and creation
operators in the Fourier plane.

We can use Eq.~(\ref{eq:quant-trans}) for calculation of the coefficients
$\hat a_k ^{(r)}$ in the reconstructed object as
\begin{equation}
     \hat a_k^{(r)} = \frac{\hat f_k}{i^k\sqrt{\lambda_k}}=\hat a_k +
     \sqrt{\frac{1-\lambda_k}{\lambda_k}} \hat b_k, %
     \label{eq:quant-rest}%
\end{equation}
where the superscript $(r)$ indicates "reconstructed". As follows from
Eq.~(\ref{eq:quant-rest}), the reconstruction of the input object is no
longer exact because of the second term in Eq.~(\ref{eq:quant-rest}). This
term contains the annihilation operators $\hat b_k$ responsible for the
vacuum fluctuations of the electromagnetic field in the area outside the
object. It is important to notice that these vacuum fluctuations prevent
from reconstruction of the higher and higher coefficients $\hat a_k$ in
the object because of the multiplicative factor
$\sqrt{(1-\lambda_k)/\lambda_k}$. Indeed, the eigenvalues $\lambda_k$
become rapidly very small after the index $k$ has attained some critical
value. This leads to rapid "amplification" of the vacuum fluctuations in
the reconstructed object that limits the number of the reconstructed
coefficients $\hat a_k$. Below we illustrate numerically the role of these
quantum fluctuations in superresolution.

\begin{figure}[h]
\begin{center}
\epsfxsize=14pc %
\rotatebox{270}{\epsfbox{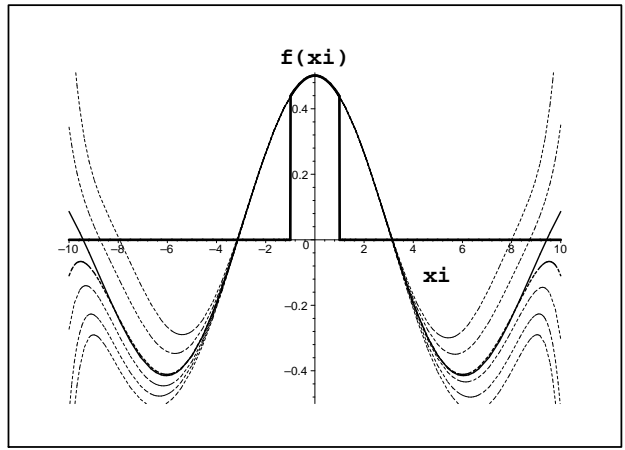}}
\end{center} %
\caption{Reconstruction of the spatial Fourier spectrum of the object
with light in a coherent state with mean photon number
$\langle N \rangle= 10^{12}$. Five dotted lines correspond to five random
Gaussian realizations of the quantum fluctuations. %
\label{fig:12-0}}
\end{figure}

The relative value of quantum fluctuations depends on the signal-to-noise
ratio in the input object which for the light in a coherent state is
determined by the total mean number of photons passed through the object
area during the observation time. For example, for a laser beam with
$\lambda=1064$ nm and optical power of 1 mW, and observation time of 1 ms
we obtain the mean photon number of $\langle N \rangle=5.3\cdot 10^{12}$.

In Fig.~\ref{fig:12-0} we have shown the results of reconstruction of the
spatial spectrum of the object from Fig.~\ref{fig:obj-img}a  when quantum
fluctuations of a coherent state are taken into account. The solid line
gives an exact spatial Fourier spectrum of the object and the solid bold
line the part of the spectrum passed through the pupil as in
Fig.~\ref{fig:spectra}. We use 6 prolate functions and the mean photon
number in the input object is taken $\langle N \rangle= 10^{12}$. The five
dotted lines correspond to five random Gaussian realizations of the
quantum fluctuations in the coherent state of $\hat a_k$ and the vacuum
fluctuations of $\hat b_k$. The dotted bold line corresponds to the
reconstructed spectrum by 6 prolate functions without noise (as in
Fig.~\ref{fig:obj-img}). One can observe that the role of quantum
fluctuations becomes more and more important as one goes to the higher and
higher spatial frequencies where the random realizations of the Fourier
spectra deviate more and more from the mean value given by the dotted bold
line.

\begin{figure}[h]
\begin{center}
\epsfxsize=14pc %
\rotatebox{270}{\epsfbox{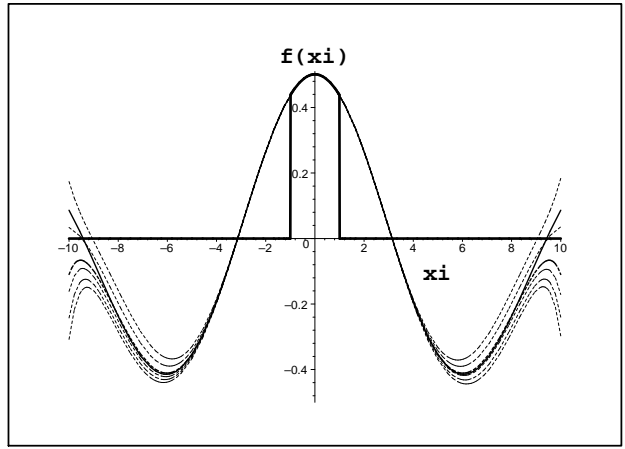}}
\end{center} %
\caption{ Same as Fig.~4 but with $\langle N \rangle= 10^{13}$.%
\label{fig:13-0}}
\end{figure}

In Fig.~\ref{fig:13-0} we have increased the total mean value of photons
to $\langle N \rangle= 10^{13}$. This corresponds to an increased
signal-to-noise ratio in the input object and should allow for better
superresolution. This is illustrated in Fig.~\ref{fig:13-0} where we can
reconstruct higher spatial frequencies in the Fourier spectrum as compared
to Fig.~\ref{fig:12-0}.

\begin{figure}[h]
\begin{center}
\epsfxsize=14pc %
\rotatebox{270}{\epsfbox{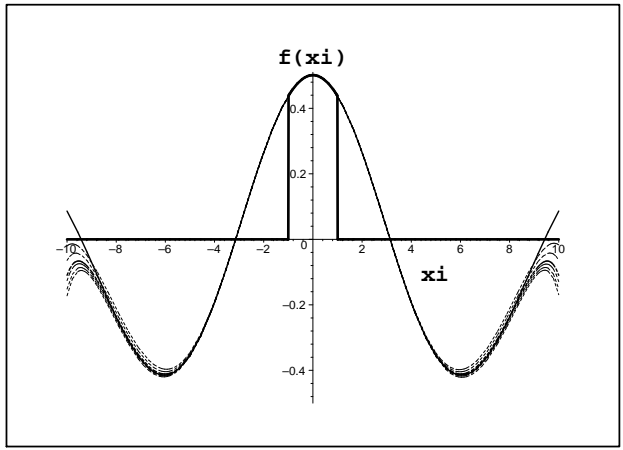}}
\end{center} %
\caption{ Reconstruction of the spatial Fourier spectrum of the object
with light in multimode squeezed state with mean photon number
$\langle N \rangle= 10^{12}$ and the squeezing parameter $e^r=10$.%
\label{fig:12-10}}
\end{figure}

The same result can be achieved by using multimode squeezed light instead
of increasing the power of the source illuminating the object. This is
illustrated in Fig.~\ref{fig:12-10} where we have used $\langle N \rangle=
10^{12}$ as in the Fig.~\ref{fig:12-0}, but considered the light in a
multimode squeezed state with the squeezing parameter $e^r=10$ instead of
the coherent state. As the result the fluctuations in the higher spatial
frequencies are decreased that gives better superresolution.

In conclusion, we have numerically investigated the role of quantum
fluctuations in reconstruction of spatial spectra of optical objects. We
have  demonstrated that when quantum fluctuations are not taken into
account one can achieve superresolution of about factor 10 over the
Rayleigh limit. We have confirmed that the limit of superresolution is set
by quantum fluctuations and depends on the signal-to-noise ratio in the
input object. For a given signal-to-noise ratio one can further improve
superresolution by using multimode squeezed light.

This work was supported by the Network QUANTIM~\cite{QUANTIM}
(IST-200-26019) of the European Union.

\end{document}